\documentclass[10pt,aps,prl,twocolumn,nofootinbib,superscriptaddress]{revtex4-2}

\usepackage{graphicx} \usepackage{psfrag} \usepackage{mathrsfs} \usepackage{amssymb, bm} 
\usepackage{amsmath, amsthm} \usepackage{epstopdf} 
\usepackage[breaklinks=true]{hyperref} \usepackage{enumerate} 
\usepackage{longtable} \usepackage{subfigure} 
\usepackage{mathrsfs} \usepackage{graphicx} 
\usepackage{natbib,url,textcase}  \usepackage{xurl} 
\usepackage[dvipsnames]{xcolor}
\usepackage[title]{appendix}
\usepackage[english]{babel} 
\usepackage[T1]{fontenc} 
\usepackage{comment} 
\usepackage{geometry}
\allowdisplaybreaks

\usepackage{tikz}

\newcommand{\be}{\begin{equation}} 
\newcommand{\ee}{\end{equation}} 

\definecolor{purple}{rgb}{1,0,1} \definecolor{lime}{HTML}{a6CE39} 

\newcommand{\orcidicon}{%
	\begin{tikzpicture}
		\draw[lime, fill=lime] (0,0) circle [radius=0.15] 
		node[white] {{\fontfamily{qag}\selectfont \tiny ID}}; 
		\draw[white, fill=white] (-0.0625,0.095) circle 
		[radius=0.007];
	\end{tikzpicture} \hspace{-2mm} }

\newcommand\orcidValerio{{\href{https://orcid.org/0000-0002-2601-1870}{\orcidicon}}}
\newcommand\orcidMax{{\href{https://orcid.org/0000-0003-0325-3911}{\orcidicon}}}

\begin{document}

\renewcommand{\theequation}{\arabic{equation}}

\title{Black hole apparent horizons become comoving with the universe}

\author{Valerio Faraoni\orcidValerio} \email[]{vfaraoni@ubishops.ca} 
\affiliation{Department of Physics \& Astronomy, Bishop's University, 
2600 College Street, Sherbrooke, Qu\'ebec, Canada J1M~1Z7}

\author{Massimiliano Rinaldi\orcidMax} \email[]{massimiliano.rinaldi@unitn.it} 
\affiliation{Department of Physics, University of Trento, Via Sommarive 
14, 38122 Povo (TN), Italy}
\affiliation{Trento Institute for Fundamental Physics and Applications 
(TIFPA) --- INFN, Via Sommarive 14, 38122 Povo (TN), Italy}

\begin{abstract}

\noindent It has been shown that static black hole event horizons cannot 
exist in time-dependent backgrounds without becoming naked singularities. 
In light of this, we are left with the question ``how does a black hole 
horizon evolve in an expanding universe?'' Here we offer a solution to 
this puzzle by showing that spherically symmetric black hole apparent 
horizons tend to become comoving, thus confirming the necessity of a 
cosmological coupling and raising intriguing possibilities for the early 
growth of black holes.

\end{abstract}
\maketitle

\medskip \noindent {\it Introduction}---There is now compelling evidence 
that supermassive black holes at the centres of galaxies have grown very 
fast at redshifts $z>6$. It is very difficult, if not impossible, to 
explain this growth via standard astrophysical channels 
\cite{Inayoshi:2019fun,Volonteri:2021sfo}. Stellar mass black holes are 
not free of problems either: the gravitational waves detected by {\em 
LIGO} and arising from binary mergers point to the existence of black 
holes in a mass gap forbidden by well-established stellar evolution 
\cite{LIGOScientific:2020iuh,KAGRA:2021vkt,Mehta:2021fgz}. Given these 
difficulties, it is worth considering whether fundamental physics is 
missing in the description. In other words, it is possible that there is 
some fundamental reason for early growth that is completely independent of 
environmental factors.

Apart from interactions with test fluids and plasmas, astrophysical black 
holes are usually modelled using the vacuum Kerr solution of general 
relativity, which is isolated and asymptotically flat \cite{Wald:1984rg},  
but  astrophysical black 
holes are embedded in the universe instead. Astrophysical and cosmological 
scales are vastly separated but, over billions of years, tiny effects of 
the cosmic expansion on black holes could accumulate and become 
astrophysically relevant.  
Ref.~\cite{Farrah:2023opk} studied black hole populations in red 
elliptical galaxies at redshifts $0<z\leq 2.5$ and reported tentative 
evidence of coupling between the masses of supermassive black holes and 
the scale factor of the ambient Friedmann-Lema\^itre-Robertson-Walker 
(FLRW) universe \cite{Farrah:2022jgp,Farrah:2023opk},  which is also in 
agreement with the recent DESI results 
on the redshift evolution of the dark energy equation of state 
\cite{DESI:2025ffm,Croker:2024jfg}. Potentially observable effects might 
also appear in the spectrum of the stochastic background of gravitational 
waves generated by slowly inspiralling supermassive black holes 
\cite{Calza:2024qxn,Lacy:2023kbb} and in the orbital  timing of black 
hole-pulsar binary systems \cite{Croker:2019mup}.
Although very preliminary and widely 
questioned 
(\cite{Rodriguez:2023gaa,Andrae:2023wge,Gao:2023keg,Amendola:2023ays, 
Lacy:2023kbb, 
Cadoni:2023lqe,Cadoni:2023lum,Cadoni:2024jxy,Croker:2021duf,Cadoni:2024rri}  
and references therein), the suggestion that
 black holes may ``feel'' the cosmic expansion, given enough time to 
sample it, cannot be discarded {\it a priori}.

A more speculative context in which black holes cannot be approximated by 
the asymptotically flat Schwarzschild or Kerr solution is that of 
primordial black holes in the early universe. At very early epochs, and 
especially during inflation, the Hubble scale is much smaller than in the 
present universe and the separation between black hole and cosmological 
scales is not as sharp  
\cite{Boehm:2020jwd,Harada:2021xze,Hutsi:2021nvs,Calza:2026csk}.

The search for exact solutions of the Einstein equations describing black 
holes (or other objects) in a FLRW universe began with McVittie 
\cite{McVittie:1933zz} and Einstein \& Straus \cite{Einstein:1945id}, but 
has not been fruitful due to the lack of generic solutions (as opposed to 
asymptotically flat black holes in which case the Kerr solution is 
generic). For this reason, here we study the problem of the evolution of 
black holes in general terms without referring to specific solutions of 
the Einstein equations.

In a previous work we showed that a black hole {\em event} horizon (a null 
surface) that locally resembles a Schwarzschild one cannot be rigorously 
static (even momentarily) in an evolving 
environment, since this requirement forces it to become a naked (null) 
singularity (\cite{Faraoni:2024ghi}, see also \cite{Davidson:2012si}). 
This     
result raises the question of how a black hole evolves in the specific 
environment of an expanding FLRW universe. Time-dependent black holes are 
best described by {\em apparent} horizons (AHs), which are defined 
quasi-locally, instead of event horizons which are teleological and can 
only be defined by knowing the entire spacetime structure, including 
future infinity.

\medskip
\noindent {\it Apparent horizons}---
To describe a central black hole embedded in a FLRW universe, consider 
the line element\footnote{We adopt the notations of 
Ref.~\cite{Wald:1984rg}.} 
\be\label{metric}
ds^{2}=-T^{2}\left( t,r \right) dt^{2} 
+ a^{2}\left( t,r \right)\left(dr^{2}+r^{2}d\Omega^{2}_{(2)}\right) 
\ee
depicting a rather general time-dependent and 
spherically symmetric geometry in coordinates that are 
simultaneously isotropic and comoving. Here, 
$d\Omega^{2}_{(2)}=d\vartheta^{2}+\sin^{2}\vartheta \, d\varphi^{2}$ is 
the 
line element  on the unit 2-sphere and $T \geq 0, a>0$. At large radial 
distances $r$ from 
the central black hole, $T(t,r) $ must reduce to unity and $a(t,r) 
\simeq 
a(t)$ loses its spatial dependence (we restrict to spatially flat FLRW 
``backgrounds'' for simplicity, 
but there is no problem of principle in considering spatially curved 
universes). 
The metric \eqref{metric} restricts our analysis to 
time-dependent, spherically symmetric spacetimes that  admit a diagonal, 
spatially conformally flat spatial slicing. This means that we are 
essentially excluding a non-vanishing 
radial flux 
through the horizon (e.g., Vaidya or Sultana-Dyer models). 
Nevertheless 
the quantities that matter, namely the Misner-Sharp-Hernandez  mass and 
the location of the apparent horizon are gauge-invariant and geometrically 
robust. 

The areal radius is the geometric invariant $R\left( t,r 
\right)=a\left( t,r \right) \, r$. If apparent horizons exist, their 
physical radii $R_\mathrm{AH}(t)= a\left( t, r_\mathrm{AH}(t) \right) 
r_\mathrm{AH}(t)$ are determined by the real positive roots of the 
equation 
\be\label{eq:ah}
\nabla^{c} R(t,r) \nabla_{c} R(t,r)=0 
\ee
(see e.g., \cite{Faraoni:2015ula}), where 
\be
\nabla_{c}R(t,r)=\delta_{c}^{0} \, \dot{a}\left( t,r \right)  
r + \delta_{c}^{1} \left[ a\left( t,r \right)+r a'\left(t,r \right) 
\right] \,.
\ee
In these expressions, and from now on, an overdot and  a prime denote, 
respectively,  partial differentiation with respect to $t$ 
and $r$. Since $\dot{R}= \dot{a} r $ and $R'= a+a'r$, at the apparent 
horizon Eq.~\eqref{eq:ah} yields
\be
\frac{\dot{a}_\mathrm{AH} r_\mathrm{AH} }{T_\mathrm{AH}} 
=\pm \left( 1+ 
\frac{a'_\mathrm{AH} r_\mathrm{AH}}{a_\mathrm{AH}}  \right) \,,
\ee
where the positive sign describes a cosmological apparent horizon since, 
for large 
radii, it 
reduces to the well-known expression  $r_\mathrm{AH} 
\simeq  1/\dot{a}$, or 
$R_\mathrm{AH}\sim 1/H$, where $H\equiv \dot{a}/a$ is the Hubble 
function \cite{Faraoni:2015ula}. Choosing the negative sign, which describes the black hole 
apparent horizon, we have 
\be\label{eq:dotR}
\dot R_\mathrm{AH}=-{T_\mathrm{AH}R'_\mathrm{AH}\over a_\mathrm{AH}}\,.
\ee
From now on we drop the arguments of the functions $r_\mathrm{AH}$ and 
$a_\mathrm{AH}$  unless necessary. 
  
In an expanding universe $\dot R=\dot a r>0$, thus we infer from Eq.\ 
\eqref{eq:dotR} that $R'_\mathrm{AH}<0$.  $\dot R=\dot a r$  by 
construction since 
$\dot r=0$, thus we also have $\dot R_\mathrm{AH}=\dot a_\mathrm{AH}  
r_\mathrm{AH}$ (however, this does not imply that $\dot r_\mathrm{AH} 
\equiv dr_\mathrm{AH}/dt $ vanishes).
With this in mind, we write Eq.\ \eqref{eq:dotR} as
\be\label{eq:aux1}
{\dot a_\mathrm{AH} r_\mathrm{AH} \over T_\mathrm{AH}} + {a'_\mathrm{AH} 
r_\mathrm{AH} \over  a_\mathrm{ AH}}+1=0\,.
\ee
In an expanding universe $\dot a>0$ everywhere,  thus also\footnote{Even 
in the McVittie solution where the black hole apparent horizon 
becomes smaller \cite{Faraoni:2012gz,Faraoni:2015ula},  $\dot{a}$ is 
positive everywhere outside the  null singularity at constant finite 
radius (including at the apparent horizon), vanishing 
only at the singularity itself.} 
$\dot a_\mathrm{ AH}>0$ and  the equation above can only be satisfied if  
$a_\mathrm{ AH}'<0$. (Since it cannot be 
$\dot{a} =  0$ \cite{Faraoni:2024ghi,Davidson:2012si}, it could be 
that $\dot{a}<0$ but  $a(t)$ is 
bounded from below by zero and then can only approach a constant value 
$a_0$ 
asymptotically as 
$t\to+\infty$, or else $a\to 0$ in a finite time---in both cases a 
physically pathological situation is approached, which excludes 
$\dot{a}<0$.)

Our goal is to understand whether the black hole apparent horizon 
expands, and at which rate. A natural question is whether the black hole 
apparent horizon can expand faster, slower, or at the same rate as the 
cosmic 
substratum in 
which it is embedded. Therefore, the main quantity to study is the 
total time derivative of the  areal radius of the apparent horizon.

By recalling that 
$ \dot{R}_\mathrm{AH}=\dot{a}_\mathrm{AH} r_\mathrm{AH}$  we find
\be\label{eq:dRdt}
{dR_\mathrm{AH}\over dt}=\dot R_\mathrm{AH}+\dot r_\mathrm{AH}
a_\mathrm{AH}\left(1+{a'_\mathrm{AH}r_\mathrm{AH} \over 
a_\mathrm{AH}}\right)\,.
\ee
By combining Eqs.\ \eqref{eq:aux1} and\ \eqref{eq:dRdt}, we find 
\be\label{eq:mass}
{1\over R_\mathrm{AH}}{dR_\mathrm{AH}\over dt}=H_\mathrm{AH}\left( 
1-{a_\mathrm{AH}\dot r_\mathrm{AH}\over T_\mathrm{AH}} \right) ={1\over 
M}{dM\over dt}\,,
\ee
where $H_\mathrm{AH}$ is the Hubble function at the apparent 
horizon. The last equality follows from the definition of the 
Misner-Sharp-Hernandez mass $M$ \cite{Misner:1964je,Hernandez:1966zia}  
\be
1-\frac{2M(t,R)}{R} = \nabla^c R \nabla_c R \,.
\ee
This quantity is universally used in studies of black 
hole 
collapse in spherical symmetry. At the apparent  
horizon, $2M_\mathrm{AH}=R_\mathrm{AH}$ (which mimics 
the 
well known expression of the Schwarzschild radius). The value of the 
quantity within the round brackets of Eq.~\eqref{eq:mass} is 
crucial 
to determine the growth rate of the black hole mass $M(t)$,  
hence the strength of the cosmological coupling.  Three possible 
situations arise  in principle, which we discuss separately. 
They do not necessarily occur, but we need to contemplate them all to figure out 
 the only possible physical situation.

\medskip
\noindent {\it Exactly comoving apparent horizon} --- In this case 
$\dot{r}_\mathrm{AH}=0$, 
$r_\mathrm{AH}=r_0=$~const., and
\be\label{eq:comAH}
R_\mathrm{AH}=a\left( t, r_0\right) \, r_0
\equiv a(t)\, r_0\,, 
\ee
\be
M(t)=M_0 \, a(t)
\ee
(with $M_0$ constant)  are exactly comoving with the 
cosmic fluid and  
\be
\frac{1}{M}\,  \frac{dM}{dt} = \frac{1}{R_\mathrm{AH}} \, 
\frac{dR_\mathrm{AH}}{dt} = H_\mathrm{AH} \,. \label{questa}
\ee

\medskip
\noindent {\it Slower than comoving apparent horizon} --- If 
$\dot{r}_\mathrm{AH}>0$, the {\it comoving} radius 
$r_\mathrm{AH}$ of the apparent horizon grows, $ 
\left(1-{a_\mathrm{AH}\dot r_\mathrm{AH}/ T_\mathrm{AH}}\right) < 1 $, 
and 
\be
\frac{1}{\tau_\mathrm{BH}} \equiv \frac{1}{R_\mathrm{AH}}  \, 
\frac{dR_\mathrm{AH}}{dt}= 
\frac{1}{M} \, \frac{dM}{dt} <H_\mathrm{AH} 
\equiv \frac{1}{\tau_\mathrm{cosmic}}  \,,
\ee
or $\tau_\mathrm{BH}> \tau_\mathrm{cosmic}$, where $\tau_\mathrm{BH, 
cosmic}$ are the time scales of variation of the black hole apparent 
horizon and of the 
surrounding cosmological ``background'', respectively. 

Let us discuss first the special case  $R_\mathrm{AH}=$const.,  
corresponding  to the critical value $\dot 
r_\mathrm{AH}=T_\mathrm{AH}/a_\mathrm{AH}$. 
Radial null 
geodesics have tangent vector $u^c =\left( u^{0}, u^{1}, 0, 
0 \right)$.  The normalization $u^c u_c=0$ imposes that $u^{1}/u^{0}=\pm 
T/a$ (positive sign for outgoing, negative sign for ingoing geodesics), 
i.e.,  the 
radial coordinate velocity of the apparent horizon coincides with the 
speed of light, hence it must be a null horizon and an event horizon.  
However, this possibility has already been  excluded in 
\cite{Faraoni:2024ghi,Davidson:2012si}. The same conclusion can be reached 
by considering 
the parametric equation of  the apparent horizon 
\be
 f\left( t,r \right) = r-r_\mathrm{AH}(t) =0 \,.
\ee
The normal vector to this  surface is $ N_{c}=\partial_{c} f|_\mathrm{AH} $ and 
its norm reads
\be
N_a N^a ={1\over a_\mathrm{AH}^{2}}-{\dot r_\mathrm{AH}^{2}\over  
{ T_\mathrm{AH}^2}}  \,.
\ee
If the apparent horizon is a null surface, then
\be
\dot r_\mathrm{AH}=\pm {T_\mathrm{AH}\over a_\mathrm{AH}} \,,
\ee
which leads to the same situation discussed above. 

Since $dR_\mathrm{AH}/dt$ cannot vanish (even instantaneously) in an 
expanding universe, the physical apparent horizon  radius $R_\mathrm{AH}$ 
always 
increases or always decreases.

If $\dot r_\mathrm{AH}$ is initially larger than the critical value, $ 
\dot r_\mathrm{AH}>T_\mathrm{AH}/a_\mathrm{AH}>0$, then $R_\mathrm{AH}$ and $M$ decrease 
monotonically.\footnote{The fact that $M$ decreases can be interpreted by 
saying that, in comparison with a comoving sphere which incorporates more 
and more cosmic material as it expands, a less-than-comoving sphere loses 
mass.} Since they are bounded from below by zero and monotonic, they 
eventually either vanish, which is unphysical for realistic black holes, 
or they asymptote to constant values and the black hole approaches the 
pathological null event horizon-singularity, which becomes barely covered 
by the apparent horizon.  By ``pathological'' we mean a solution 
that hosts a null, naked singularity at the would-be static event horizon, 
according to what was shown in \cite{Faraoni:2024ghi}.  The well-known 
McVittie solution \cite{McVittie:1933zz} is a concrete example of this 
scenario \cite{Faraoni:2012gz,Faraoni:2015ula}. Moreover, by slowing down 
its evolution, the black hole 
horizon  
eventually 
reaches an adiabatic regime in which black hole thermodynamics is 
meaningful \cite{Hayward:1997jp, 
Hayward:1998ee,Hayward:2008jq,Faraoni:2015ula} and the decreasing apparent 
horizon  area 
${\cal A}_\mathrm{AH}=4\pi R^2_\mathrm{AH}$ implies a decreasing entropy 
$S={\cal A}_\mathrm{AH}/4$, contradicting a tenet of black hole physics 
(cf. Ref.~\cite{Faraoni:2024ghi}).

If $\dot r_\mathrm{AH}$ is initially smaller than the critical value, $0< 
\dot{r}_\mathrm{AH}< T_\mathrm{AH}/a_\mathrm{AH}$, then $R_\mathrm{AH}$ 
and $M$ increase monotonically, either diverging or approaching constant 
values from below. In the second case, they approach asymptotically a 
forbidden state. The black hole would then be contained {\it inside} a 
naked null singularity dividing spacetime into two disconnected regions, 
which is completely unphysical.

\begin{figure}[t]
    \centering
    \includegraphics[width=\columnwidth]{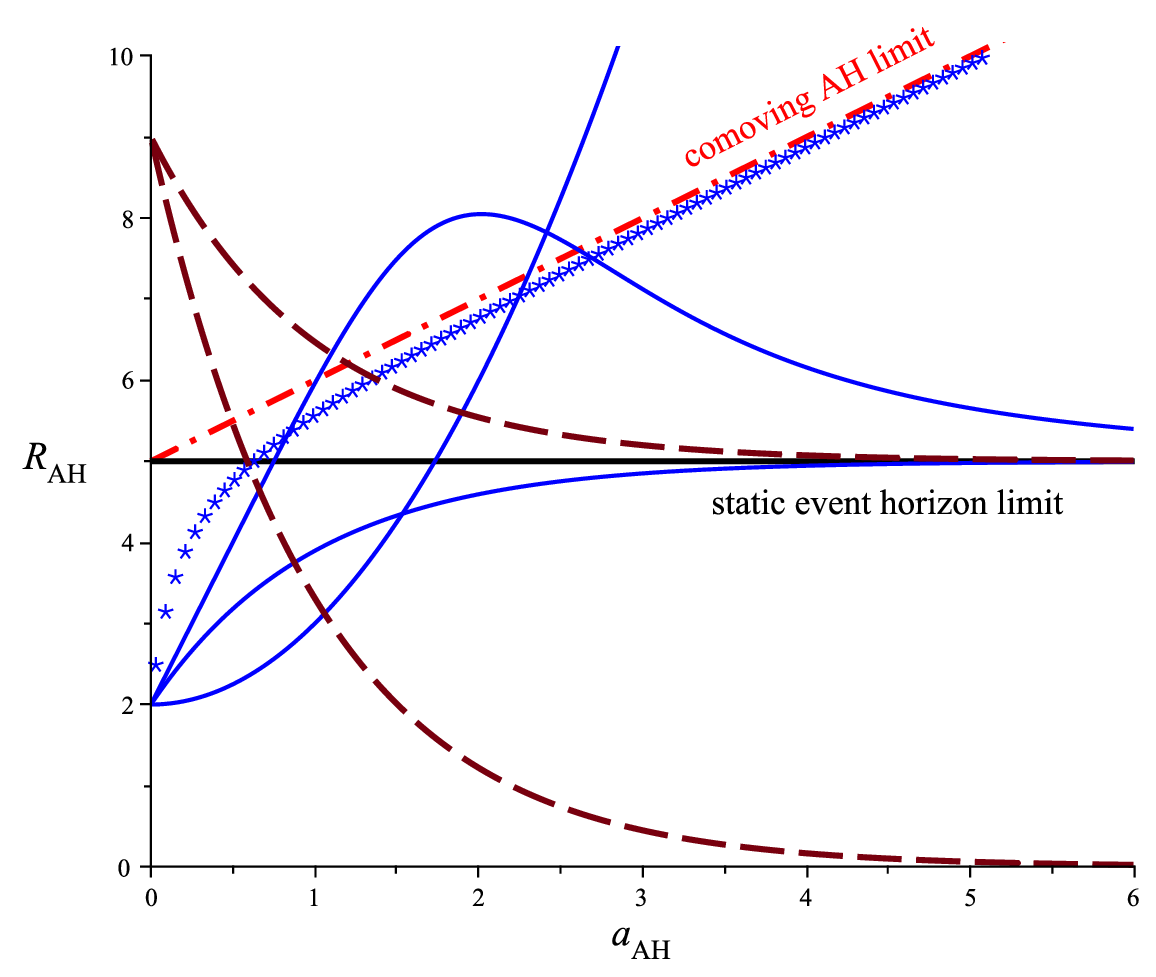}
    \caption{Qualitative evolution of slower than comoving apparent 
horizons. The plot 
    depicts $R_{\rm AH}$ vs. $a_{\rm AH}$. The tilted dot-dashed (red) 
straight line is the comoving apparent horizons described by Eq.\ 
\eqref{eq:comAH}. 
The  
horizontal black line represents the  forbidden configuration of 
a static event horizon. The solid (blue) curves  represent the possible 
evolution of apparent horizons with $0<\dot r_{\rm AH}<T_{\rm AH}/a_{\rm 
AH}$.  The 
special curve marked with asterisks asymptotes to a comoving apparent 
horizon, that  is 
$\dot r_{\rm AH}\rightarrow 0^{+}$. Finally, the dashed (brown) curves  
represent the possible cases with $\dot r_{\rm AH}>T_{\rm AH}/a_{\rm AH}>0$.}
        \label{fig1}
\end{figure}

In the first case, since the  universe expands and $ a_\mathrm{AH}\to 
+\infty $ as $t\to +\infty$, there are two possibilities. Either 
$\dot{r}_\mathrm{AH} \to 0$, in which case the black hole apparent horizon  
becomes 
comoving, or else $\dot{r}_\mathrm{AH}$ remains finite and positive with 
$ a_\mathrm{AH}\dot{r}_\mathrm{AH}/T_\mathrm{AH} \to +\infty$. Then, at 
some point the second 
term in the round bracket of Eq.\ \eqref{eq:mass}  comes to dominate and  
the expansion rate of the apparent horizon radius can be approximated by
\be\label{eq:largea}
{1\over M} \, {dM\over dt} \simeq -H_\mathrm{AH} \, {a_\mathrm{AH} \, 
\dot r_\mathrm{AH} \over T_\mathrm{AH}} <0 \,.
\ee
Thus, $M$ and $R_\mathrm{AH}$ decrease monotonically in time asymptoting 
to constant values and approaching the forbidden solution 
$R_\mathrm{AH}=$~const., which is again unphysical.\footnote{ This 
is 
the case of the McVittie solution \cite{McVittie:1933zz}, in which the 
would-be event horizon of the 
central Schwarzschild black hole, when embedded in a FLRW universe, 
becomes a null singularity at constant areal radius covered by the 
apparent horizon which asymptotes to it 
\cite{Faraoni:2015ula,Faraoni:2012gz}.}  
In Fig.~\ref{fig1} we qualitatively show  the possible evolutions of 
slower than comoving apparent horizons.

\medskip
\noindent {\it Faster than comoving $R_\mathrm{AH}$} --- The last 
possibility is $\dot{r}_\mathrm{AH}<0$, corresponding to $
\left(1-{a_\mathrm{AH}\dot r_\mathrm{AH}/ T_\mathrm{AH}}\right) >1 
$ in Eq.\ \eqref{eq:mass} and to 
\be
\frac{1}{\tau_\mathrm{BH}}=  \frac{1}{M} \, 
\frac{dM}{dt}=\frac{1}{R_\mathrm{AH}} \, 
\frac{dR_\mathrm{AH}}{dt}> H_\mathrm{AH} =\frac{1}{\tau_\mathrm{cosmic}}   
>0 \,,
\ee
or $\tau_\mathrm{BH} < \tau_\mathrm{cosmic}$. $M$ and $R_\mathrm{AH}$ 
always increase (if $H\geq 0$) and either asymptote to constant values 
from below or diverge. In the first case, they would be contained inside a naked null 
singularity dividing spacetime into two disconnected regions, which is 
unphysical. In the second case, since $r_\mathrm{AH}$ decreases 
and is bounded from below by zero, it either reaches zero with non-zero 
derivative and the apparent horizon disappears, which is unphysical, or 
else it 
asymptotes to a constant value $r_0>0$, with $M$ and $R_\mathrm{AH}$ 
becoming asymptotically comoving.

To summarize, the only physically viable scenario is one where $\dot 
r_\mathrm{AH}\rightarrow 0$, which implies that once an apparent horizon 
has formed, it is destined to become comoving because of its intrinsic 
coupling to the expanding universe.

\medskip
\noindent {\it Conclusions} --- For simplicity, we have limited our 
discussion to spherically symmetric 
black holes while realistic ones rotate, but the fundamental result that 
black holes adjust their apparent horizons to the cosmic expansion 
by becoming comoving with the 
surrounding FLRW universe is expected to apply also to rotating objects, 
which will be studied in future work. A fundamental limitation of our 
discussion is the characterization of time-dependent black holes by their 
apparent horizons, which are foliation-dependent. However, quasilocal 
apparent horizons and marginally trapped surfaces are the tools used 
universally in situations where black holes evolve in time, such as in the 
black hole mergers originating the gravitational waves studied by {\it 
LIGO}, {\it VIRGO}, and {\it KAGRA}, where the teleological event horizon 
is useless \cite{LIGOScientific:2020iuh,KAGRA:2021vkt}.  Moreover, in 
spherical symmetry all spherically symmetric foliations agree on the 
location and structure of the apparent horizon \cite{Faraoni:2016xgy}. 

A reasonable concern could be that also comoving apparent 
horizons develop a naked singularity, analogously to static event horizons 
\cite{Faraoni:2024ghi}. However, in a recent work, Cadoni et al. have 
found a class of exact solutions of the Einstein equations describing 
cosmologically coupled black holes with comoving and non-singular apparent 
horizon \cite{Cadoni:2026ejk}. These solutions are consistent with our 
analysis and demonstrate the possibility that asymptotically 
comoving apparent horizons are indeed singularity-free. 

The previous considerations did not use the Einstein (or other) 
field equations, but only geometry and reasonable physical assumptions. 
We have neglected the astrophysical environment to which black holes are 
unavoidably coupled but, given the difficulties in explaining the fast 
growth 
of supermassive black holes via standard astrophysical channels such as 
accretion, mergers, or the interaction with their host galaxies, it makes 
sense to ask whether their early growth may be due to fundamental physics 
instead. Answering this question requires abstraction from ``accidental'' 
astrophysical processes (and, indeed, their effects would be minimized in 
red elliptical galaxies which have been inactive 
for  a long time \cite{Farrah:2023opk}). To study this problem of 
principle, we have assumed a FLRW description down to  spatial 
scales smaller than the ones over which spatial averages motivate this 
geometry in cosmology, in line with  \cite{Croker:2022vdq}.

The main conclusion to this problem of principle, which may come as a 
surprise, is that black holes embedded in the universe eventually adjust 
themselves to become comoving with it,  independently of the velocity of the horizon at its creation. The determination of the specific 
time scales for this fundamental process, and its consequences for 
astrophysics, will necessarily require specific scenarios, but this 
conclusion is already far-reaching. Our result goes in the right direction 
to solve the puzzle of how can supermassive black holes grow so big so 
quickly at early epochs; however, we make no claim, which would be 
premature and can only be justified by detailed analyses.

On typical astrophysical time scales much smaller than the Hubble time, 
the fact that black holes are really time-dependent and eventually 
comoving is, of course, irrelevant. However, black holes exist for 
substantial fractions of the age of the universe and tiny fundamental 
effects could accumulate and add to astrophysical ones (as suggested in 
\cite{Farrah:2023opk}).

A popular parametrisation for the cosmological coupled 
mass of a black hole implies that it grows as a power $k$ of the scale 
factor \cite{Croker:2021duf}. The comoving case corresponds to $k=1$, 
while $k=3$ is interesting as it can mimic a dark energy-dominated 
universe \cite{Croker:2019mup}. Our findings do not exclude the latter 
possibility, since the actual coupling function, which is 
model-dependent, might include a transient phase with $k=3$. 
In particular, the result obtained here opens up interesting possibilities 
concerning primordial black holes formed during inflation, where the 
Hubble time scale is very short. Since the scale factor $a(t)$ grows 
nearly exponentially during inflation, $\dot r_\mathrm{AH}$ must decrease 
at the same pace to keep the apparent horizon  away from a static 
configuration. As a 
result, from Eq.\ \eqref{questa} we infer that the growth rate of the 
black hole mass can be very large,  suggesting a possible, fundamental,  
growth mechanism forming larger black holes than expected during the 
early universe.

The question whether our result applies also to more general geometries 
with non-vanishing radial flux through the horizon is very intricate and 
will be hopefully addressed in a future work.

\medskip \noindent We thank K. Croker, L. Gallerani, and V. Toth for 
useful feedback on an earlier version of this work. V.F. is supported, in 
part, by the Natural Sciences \& Engineering Research Council of Canada 
(Grant No. 2023-03234).

\end{document}